\documentclass[a4paper,12pt,twoside]{article}
\usepackage[utf8]{inputenc}
\usepackage[T1]{fontenc}
\usepackage[english]{babel}
\usepackage{fancyhdr}
\usepackage{graphicx,url,color}
\usepackage{a4wide}
\usepackage{lineno}
\usepackage{authblk}
\usepackage{setspace}
\usepackage[document]{ragged2e}
\usepackage{natbib}

\begin{document}
\title{Could sampling make hares eat lynxes?}
\date{}
\author[1,2]{Brenno Caetano Troca Cabella \thanks{cabellab@gmail.com}} 
\author[2]{Fernando Meloni \thanks{fernandomeloni@usp.br}}
\author[2,3]{Alexandre Souto Martinez \thanks{asmartinez@usp.br}}
\affil[1]{Sapra Assessoria \\ Cid Silva C\'{e}sar, 600, 13562-400 - S\~{a}o Carlos - S\~{a}o Paulo, Brazil}
\affil[2]{Faculdade de Filosofia, Ci\^{e}ncias e Letras de Ribeir\~{a}o Preto (FFCLRP) Universidade of S\~{a}o Paulo (USP) Av. Bandeirantes 3900, 14040-901 Ribeir\~{a}o Preto, SP, Brazil}
\affil[3]{Instituto Nacional de Ci\^{e}ncia e Tecnologia em Sistemas Complexos (INCT\-SC/CNPq) Av. Bandeirantes 3900, 14040-901 Ribeir\~{a}o Preto, SP, Brazil}

\maketitle

\thispagestyle{fancy} 


\newpage
\begin{abstract}
Cycles in population dynamics are widely found in nature. 
These cycles are understood as emerging from the interaction between two or more coupled species. 
Here, we argue that data regarding population dynamics are prone to misinterpretation when sampling is conducted at a slow rate compared to the population cycle period. 
This effect, known as aliasing, is well described in other areas, such as signal processing and computer graphics. 
However, to the best of our knowledge, aliasing has never been addressed in the population dynamics context or in coupled oscillatory systems. 
To illustrate aliasing, the Lotka-Volterra model oscillatory regime is numerically sampled, creating prey-predator cycles. 
Inadequate sampling periods produce inversions in the cause-effect relationship and an increase in cycle period, as reported in the well-known hare-lynx paradox. 
More generally, slow acquisition rates may distort data, producing deceptive patterns and eventually leading to data misinterpretation.
\end{abstract}
\newpage

\section*{Introduction}
Quantitative sampling provides the most important information source for ecological modeling. 
Because the validation of sampling methods is a difficult issue, the results of theoretical models are rarely achieved in \textit{in situ} experiments. 
For instance, predictions concerning scaling and survival/extinction transition in population dynamics~\citep{Cabella2011, Cabella2012, Santos2014} have not been experimentally tested.
Consequently, many ecological data may lack concordance, and/or the real systems may be profoundly misinterpreted.

An important example, and not yet fully understood, lies in the periodic species abundance cycles in population dynamics. 
These cycles may appear in coupled systems, in which two or more species interact due to a cause-effect relationship. 
Using the historic data series from Hudson’s Bay Company, MacLulich~\citep{MacLulich1937} and Elton and Nicholson~\citep{EltonNicholson1942} found regular cycles in the population of Snowshoe Hares (\textit{Lepus americanus}) and Canadian Lynx (\textit{Lynx canadensis}). 
Data of both of these species were matched and indicated an overlap with a small delay between the species abundance. 
The system was interpreted from the perspective of trophic interactions, as a regular predator-prey system, which was first labeled the Lotka-Volterra model (LVM)~\citep{Odum1953}. 
Some years later, the model became more robust, considering finite limits in the oscillatory predation rate~\citep{RosenzweigMacArthur1963}.

Although predator-prey models are intuitively coherent and produce qualitative patterns found in nature, such models provide poor adjustment to the field data, so their empiricism is still controversial~\citep{Murray2002}. 
In "Do hares eat lynx?"~\citep{Gilpin1973}, the author fitted real data using different coupled models, and beyond the poor fit, data displayed a cause-effect relationship inversion. 
Hares seemed to negatively affect the lynx population. 
The proposed solution to this paradox was the human influence on data collection, a plausible but untestable hypothesis~\citep{Murray2002, Gilpin1973}. 
Further, field data infer that hares and lynx present regular population cycles with approximately ten years between the respective peaks, instead of the expected one-year period. 
However, neither several years of field research nor theoretical approaches could identify which factors influence this period increase~\citep{Krebsetal2001}. 

Experimental evidence has shown that both the predator and prey densities affect the dynamics of hares. 
Conversely, space, food sources, diseases and parasites are variables that are neglected by the models and have been experimentally discarded as modulators of cycles in hare populations~\citep{ Krebsetal2001, Smith1983, SovellHolmes1996, Krebsetal1985, Krebsetal1995}.
Therefore, extensive research only reinforces the empiric value of cyclic models as the descriptor of the predator-prey dynamics, while the poor fit is the main argument against the use of such models.

We argue that the cyclic dynamics are particularly influenced by the sampling rate. 
Apparent inversion of cycle direction or an increase in the cycle period, among other behaviors, may be artifacts due to poor sampling. 
Here, we conjecture that the aliasing effect could be a plausible explanation for the lack of concordance between oscillatory theoretical models and field data in Ecology. 
Further, aliasing shows a predictive character, which allows one to avoid possible misinterpretations when sampling is the basis for modeling. 

\section*{Material and methods} 
Before describing the methods and numerical experiment, we first introduce the necessary theoretical background. 
We present the temporal aliasing effect and the Lotka-Volterra model. 
Next, we numerically solve the model and sample it with different rates.

\subsection*{Aliasing Effect in the Lotka-Volterra Model} 
\label{sec:aliasing}

A temporal aliasing effect occurs when the sampling rate is not fast enough compared to the system’s natural cycle period. 
For example, in movies, the spiked wheels on horse-drawn wagons sometimes appear to turn backwards, the "wagon-wheel effect", which is depicted in Fig.~1. 
A wheel indeed turns clockwise, but due to the slow sampling by the camera (number of frames per second), a filmed wheel appears to turn counter-clockwise. 
This effect can be avoided considering the Nyquist-Shannon sampling theorem, which states that given a time series with minimum period $\tau_{e}$, the equally spaced intervals between samples $T_{s}$  must be smaller than half the minimum period, {\it i.e.}, $T_{s} < \left(\tau_{e}/2\right)$.

\begin{figure}[!h]
\begin{centering}
	\includegraphics[width=\columnwidth]{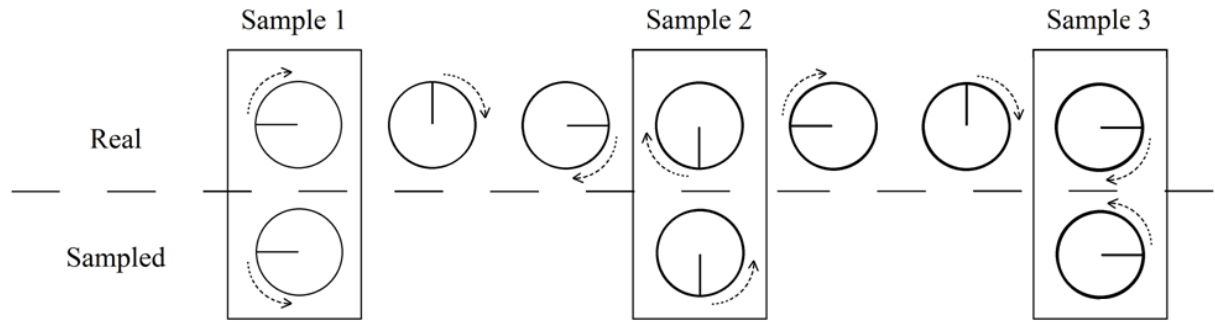}
	\caption{Example of an aliasing effect in the clockwise rotation of a wheel. 
	The visualized behavior on film is a counter-clockwise rotation, known as the wagon-wheel effect. 
	The long time interval between samples explains this curiosity.}
	\end{centering}
\end{figure}

In Ecology, cycles are extensively found in systems in which species interact with each other and with the environment ~\citep{Murray2002}. 
To illustrate the manner in which the aliasing effect may disturb the interpretation of population abundance cycles, consider a simple prey-predator interaction described by the LVM: $dx/dt=x(a-by)$ and $dy/dt=y(cx-d)$ where $x(t)$ and $y(t)$ are the prey and predator population densities, respectively, at time $t$, $a$ is the prey growth rate in the absence of predators, $d$ is the predator death rate in the absence of prey, and $b$ and $c$ are related to the interaction strength between both of the species. 
The LVM equations have two fixed points: the mutual extinction, $E_1(x^*,y^*) = (0,0)$, and the neutral center, $E_2(x^*,y^*) = (c/d,a/b)$. 
Solutions around the singular point $E_2$ are cycles with period $\tau_{e}=2\pi/\sqrt{ad}$. 
Although the LVM is not completely adequate to quantitatively describe real-world community dynamics, it is suitable here because it implies a cause-effect relation, where the number of predators increases after the prey abundance increases.

\subsection*{Numerical simulation and sampling rates}
\label{sec:numerical_simulation}

To demonstrate how sampling rates can shift the patterns in predator-prey systems, we have numerically calculated a cyclic dynamic pattern using the LVM. 
The Lotka-Volterra differential equations have been implemented in the MatLab\textsuperscript{\textregistered} language, and their solutions have been obtained using the Dormand-Prince method. 
Fig.~2~a shows the prey (full line) and predator (dashed line) population cycles. 
The model parameters have been set to produce a unitary oscillation period $\tau_{e}=1$. 

Next, the prey (circles) and predators (triangles) were sampled within fixed time intervals, $T_s$. 
We repeated the procedure, reducing the sampling rate from $\tau_{e}/5$ until $\tau_e$. 
For each sampling rate, we interpolated the points to build the respective time series to infer the original series. 
Based on the peaks of the time series, we inferred the oscillation period and the dephasing of predator and prey abundances. 
In all the cases, we considered all the individuals from both of the populations. 
Therefore, we avoided any influence of space or sampling deviation on population densities to only address the effect of sampling rate on population dynamics.

\section*{Results}
\label{sec:resultado}

In the following, we present the results of the sampling of two coupled oscillating systems. 
The main result is that the sampling rate influenced the retrieval of the original time series. 
For $T_s < \tau_e/2 = 1/2$, the system real cycle period is correctly retrieved (Nyquist-Shannon theorem), as displayed in Fig.~2b, with $T_s = \tau_e/5$.  
The fraction $\tau_e/5$ means that there were 5 sampling periods $T_s$ within $\tau_e$. 
As $T_s$ increases, the signal retrieval is increasingly biased. 
In Fig.~2c, $T_s=\tau_e/2$ is the limiting period from which the original signal can be properly retrieved. 
However, because there are only 2 sampling periods in $\tau_e$, the interpolation between the periods produces a straight line. 
Therefore, the relative delay between prey and predator dynamics cannot be correctly retrieved, and the populations seem to overlap.

For a slightly greater value, $T_s=51\tau_e/100$, different patterns can be seen in the same time series. 
Fig.~2d shows interspersed periods of synchronicity and desynchronicity. 
For $T_s=\sqrt{3/10}\tau_e$, an irrational number, the time series depicted in Fig.~2e seems to be erratic, with no identified pattern because no integer sampling periods can fit in $\tau_e$. 

A further increase in $T_s$ causes an inversion of the prey-predator cycles and an enhancement of the population cycle period. 
In Fig.~2f, $T_s=9\tau_e/10$, the predator abundance increases before the prey abundance, and when the prey abundance increases, the number of predators diminishes. 

The inverted cycle oscillations persist for even greater values of  $T_s $ as the oscillation period increases to $T_s \rightarrow \tau_e$. 
When $T_s=\tau_e$, there are no oscillations, as depicted in Fig.~2g. 
All the time series presented from Fig.~2b to Fig.~2g repeat for $k \tau_e<T_s<(2 k+1)\tau_e/2$ where $k={0,1,2,...}$.

\begin{figure}[!h]
\begin{centering}
	\includegraphics[width=0.75\columnwidth]{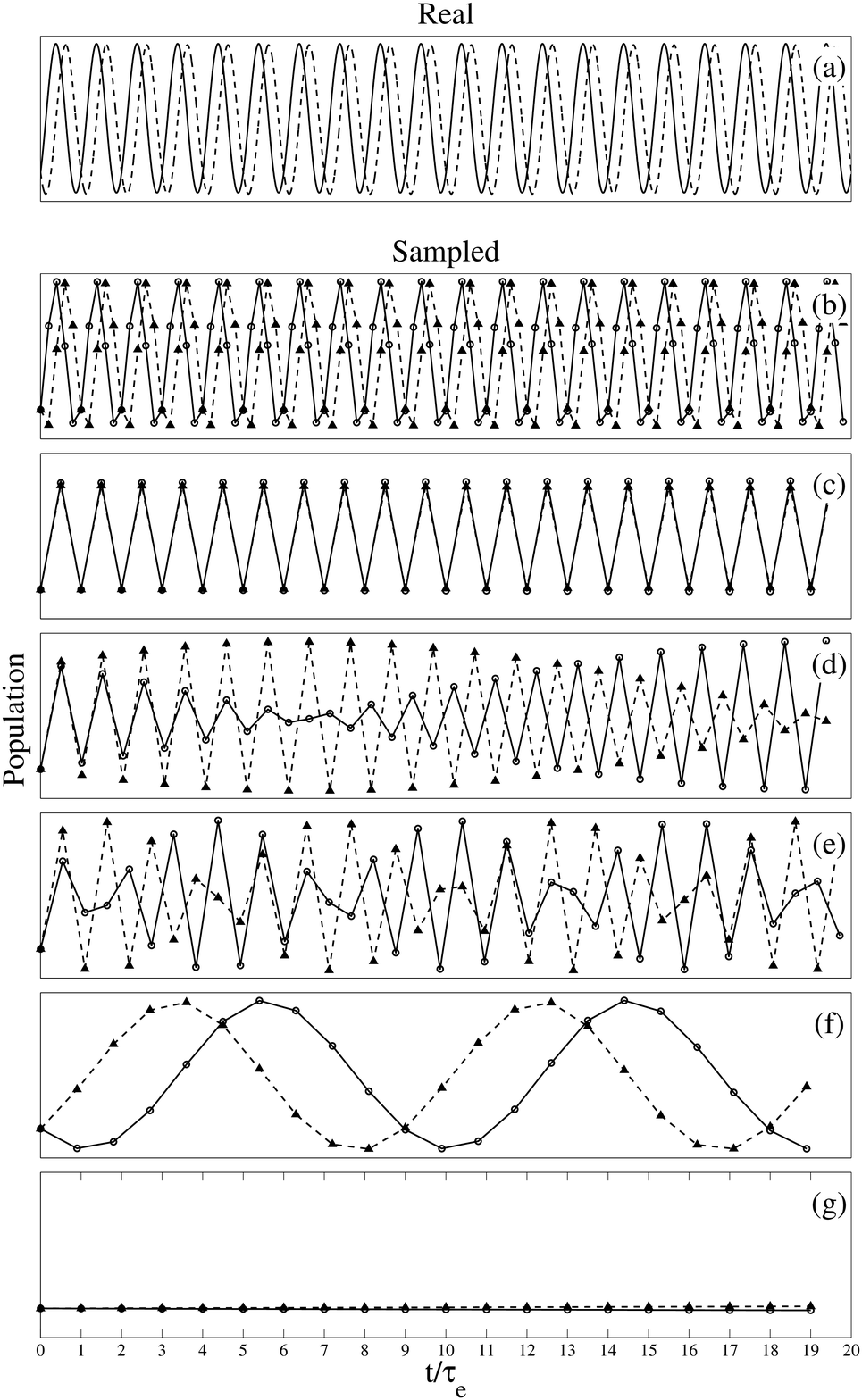}
	\caption{Prey (full line) and predator (dashed line) population cycles obtained with the Lotka-Volterra model (LVM), with oscillation period $\tau_e =1$.
	The LVM dynamics can generate different  patterns from {\bf (b)} to {\bf (g)} due only to sampling effects.
   In the following panels, prey and predator abundances are represented by an empty circle and a full triangle, respectively.
	 {\bf (b)} $T_s=\tau_e/5$, the time series correctly retrieve the LVM behavior.
	 {\bf(c)}  $T_s=\tau_e/2$,  oscillations seem to be synchronized. 
	 {\bf (d)} $T_s=51\tau_e/100$, synchronous and asynchronous patterns are present in the same series.
   {\bf (e)} $T_s=\sqrt{3/10}$, an erratic behavior emerges.
   {\bf (f)}  $T_s=9/10$, an inversion and an extension of cycle period may be interpreted as preys eating predators.
	 {\bf (g)} As $T_s \rightarrow \tau_e$, the abundances appear to be constant.}
	\end{centering}
\end{figure}

\section*{Discussion}
\label{sec:discussão}

There is a scientific consensus about field experiments that better samples lead to better interpretation, or inference, about the real pattern. 
However, the effects caused by inappropriate sampling are not trivial to analyze. 
This difficulty has already been addressed in the spatial influence on population dynamics or by the numerical insufficiency of samples~\citep{Keryetal2009, SaintJanin2014}. 
A very simple and controlled oscillatory behavior, such the one LVM simulates, may produce different time series due only to inappropriate sampling rates, Fig.~2b to Fig.~2g.
In Fig.~2f, the data series suggests that prey animals are eating predators and their cycle period is almost ten times the real one.
In the hare/lynx paradox, assuming that hunting occurs approximately every 12 months ($T_s$) and that the hare/lynx period cycle ($\tau_e$) fluctuates around this value, the aliasing effect would occur because  $T_s \approx \tau_e$. 
Therefore, it is possible that the inversion and enhancement of the prey-predator cycles may be due to sampling artifacts.
In real world systems, this difficulty is amplified because the populations’ periodicity is not necessarily constant and/or many species interactions may tangle the dynamics even more.
These results indicates that coupled systems (such as ecological systems) seem to be very sensitive to temporal aliasing. 

Regarding the practical implications of sampling rates, a paradox emerges from field studies. 
The appropriated sampling rate always depends on \textit{ad hoc} information about the real period of a species cycle. 
However, this knowledge is generally obtained by sampling the species, creating a redundant uncertainty. 
This problem could be the case of the hare-lynx system, for which almost all of the studies have used few data sources, that often were acquired from circumstantial sampling, without an adequate experimental planning

The aliasing effect is not restricted to biological experiments; it is a statistical phenomenon, and therefore, we highlight the large scope of our finding. 
A search in the scientific literature demonstrates that the aliasing effect is poorly explored, and its consideration may have deep implications. 
For instance, delays in coupled systems are ordinarily interpreted as competition effects~\citep{Vandermeer2004}, but here, we have demonstrated that these delays can also emerge from inappropriate sampling. 
Benicà and collaborators~\citep{Benincaetal2008, Benincaetal2009} have studied a long time series of plankton communities, applying regular samples to measure several species. 
The authors have found that the cause-effect relationship suggests a chaotic food web. 
Although aliasing could provide an alternative explanation to the plankton community food web, this hypothesis was not tested.

Aliasing should be better evaluated in many other circumstances, such as the coupled aerosol-cloud-rain system, because the LVM is applied to modeling~\citep{Koren2011}. 
The influence of climate anomalies has been investigated as a driver of periods in population dynamics, as in the hare-lynx system~\citep{Stenseth2007, Yanetal2013}. 
In this case, the poor fit explanation could be related to aliasing, but again, this hypothesis has not been tested yet. 
In applicable areas, species abundance rates are the basis for evaluation of biological control success in crops, and in such cases, aliasing can have great financial consequences~\citep{Snyderetal2005}. 
Sampling effects also have implications for biological conservation and species management, as in marine ecosystems, where population levels are used as a criterion to regulate fishing~\citep{Hunsickeretal2011}. 
Further, some theoretical approaches about the trade-offs in Ecology and Evolution also concern predator-prey systems, trophic interactions or population cycles, so aliasing should be addressed~\citep{Barbosaetal2005,WeitzLevin2006, Kishidaetal2010,Cortez2011, Schneideretal2012, Kalinkatetal2013}.

To conclude, we have stressed the importance of the aliasing effect in retrieving the behavior of oscillatory dynamics, for instance, in a coupled system. 
We have numerically demonstrated that slow sampling rates of this oscillatory regime, compared to the real cycle period, may lead to data misinterpretation, even when other influences are avoided. We have qualitatively compared our results with the hares/lynxes paradox and presented a new approach to this classic problem. 
We highlight the wide scope of the aliasing effect on oscillatory coupled systems and its influence on the interpretation of real-world patterns. 
The temporal aliasing hypothesis shows a predictive character and can provide new insights to old problems in Ecology and Biology. 
This effect should be considered in future experimental designs involving population dynamics in time series.

\section*{Acknowledgments}
\begin{small}
BCTC thanks the CNPq (127151/2012-5), FM thanks FAPESP (2013/06196-4), and ASM thanks the CNPq (305738/2010-0 and 485155-2013-3). 
We would like to acknowledge the organizers of the Summer Course on Mathematical Methods in Population Biology, Roberto André Kraenkel and Paulo Inácio de Knegt López de Prado, where the hare-lynx paradox was first presented to us. 
Special thanks to C. A. S. Terçariol for calling our attention to the irrational sampling period effect. Thanks to Cristiano R. F. Granzotti, Olavo H. Menin and Tiago J. Arruda for fruitful comments on the manuscript.
\end{small}

\bibliographystyle{plain}

\begin{thebibliography}{30}
\providecommand{\natexlab}[1]{#1}

\bibitem[{Barbosa et~al.(2005)Barbosa, Caldas, and Riechert}]{Barbosaetal2005}
Barbosa, P., A.~Caldas, and S.~A. Riechert. 2005.
\newblock Ecology of Predator-Prey Interactions, chap. Species Abundance
  Distribution and Predator-Prey Interactions: Theoretical and Applied
  Consequences, pages 344--368.
\newblock Oxford University Press.

\bibitem[{Beninc\`{a} et~al.(2008)Beninc\`{a}, Huisman, Heerkloss, J\"{o}hnk,
  Branco, Nes, Scheffer, and Ellner}]{Benincaetal2008}
Beninc\`{a}, E., J.~Huisman, R.~Heerkloss, K.~D. J\"{o}hnk, P.~Branco, E.~H.~V.
  Nes, M.~Scheffer, and S.~P. Ellner. 2008.
\newblock Chaos in a long-term experiment with a plankton community.
\newblock Nature 451:822--825.

\bibitem[{Beninc\`{a} et~al.(2009)Beninc\`{a}, J\"{o}hnk, Heerkloss, and
  Huisman}]{Benincaetal2009}
Beninc\`{a}, E., K.~D. J\"{o}hnk, R.~Heerkloss, and J.~Huisman. 2009.
\newblock {Coupled predator-prey oscillations in a chaotic food web.}
\newblock Ecol. Lett. 12:1367--1378.

\bibitem[{Cabella et~al.(2011)Cabella, Ribeiro, and Martinez}]{Cabella2011}
Cabella, B. C.~T., F.~Ribeiro, and A.~S. Martinez. 2011.
\newblock {Data collapse, scaling functions, and analytical solutions of
  generalized growth models}.
\newblock Phys. Rev. E 83:061902(1--7).

\bibitem[{Cabella et~al.(2012)Cabella, Ribeiro, and Martinez}]{Cabella2012}
---{}---{}---. 2012.
\newblock Effective carrying capacity and analytical solution of a particular
  case of the richards-like two-species population dynamics model.
\newblock Physica A 391:1281--1286.

\bibitem[{Cortez(2011)}]{Cortez2011}
Cortez, M.~H. 2011.
\newblock Comparing the qualitatively different effects rapidly evolving and
  rapidly induced defences on predator-prey interactions.
\newblock Ecol. Lett. 14:202--209.

\bibitem[{Elton and Nicholson(1942)}]{EltonNicholson1942}
Elton, C.~S., and M.~Nicholson. 1942.
\newblock The ten year cycle in numbers of lynx in canada.
\newblock J. Anim. Ecol. 11:215--244.

\bibitem[{Gilpin(1973)}]{Gilpin1973}
Gilpin, M.~E. 1973.
\newblock {Do Hares Eat Lynx?}
\newblock Am. Nat. 107:727--730.

\bibitem[{Hunsicker et~al.(2011)Hunsicker, Ciannelli, Bailey, Buckel, {Wilson
  White}, Link, Essington, Gaichas, Anderson, Brodeur, Chan, Chen, Englund,
  Frank, Freitas, , Hixon, Hurst, Johnson, Kitchell, Reese, Rose, Sjodin,
  Sydeman, van~der Veer, Vollset, and Zador}]{Hunsickeretal2011}
Hunsicker, M.~E., L.~Ciannelli, K.~M. Bailey, J.~A. Buckel, J.~{Wilson White},
  J.~S. Link, T.~E. Essington, S.~Gaichas, T.~W. Anderson, R.~D. Brodeur, K.-S.
  Chan, K.~Chen, G.~Englund, K.~T. Frank, V.~Freitas, , M.~A. Hixon, T.~Hurst,
  D.~W. Johnson, J.~F. Kitchell, D.~Reese, G.~A. Rose, H.~Sjodin, W.~J.
  Sydeman, H.~W. van~der Veer, K.~Vollset, and S.~Zador. 2011.
\newblock {Functional responses and scaling in predator-prey interactions of
  marine fishes: contemporary issues and emerging concepts.}
\newblock Ecol. Lett. 14:1288--99.

\bibitem[{Kalinkat et~al.(2013)Kalinkat, Schneider, Digel, Guill, Rall, and
  Brose}]{Kalinkatetal2013}
Kalinkat, G., F.~D. Schneider, C.~Digel, C.~Guill, C.~B. Rall, and U.~Brose.
  2013.
\newblock {Body masses, functional responses and predator-prey stability}.
\newblock Ecol. Lett. 16:1126--34.

\bibitem[{Kery et~al.(2009)Kery, Dorazio, Soldaat, van Strien, Zuiderwijk, and
  Royle}]{Keryetal2009}
Kery, M., R.~M. Dorazio, L.~Soldaat, A.~van Strien, A.~Zuiderwijk, and J.~A.
  Royle. 2009.
\newblock Trend estimation in populations with imperfect detection.
\newblock J. Appl. Ecol. 46:1163--1172.

\bibitem[{Kishida et~al.(2009)Kishida, Trussell, Mougi, and
  Nishimura}]{Kishidaetal2010}
Kishida, O., G.~C. Trussell, A.~Mougi, and K.~Nishimura. 2009.
\newblock {Evolutionary ecology of inducible morphological plasticity in
  predator-prey interaction: toward the practical links with population
  ecology}.
\newblock Popul. Ecol. 52:37--46.

\bibitem[{Koren and Feingold(2011)}]{Koren2011}
Koren, I., and G.~Feingold. 2011.
\newblock {Aerosol-cloud-precipitation system as a predator-prey problem.}
\newblock P. Natl. Acad. Sci. USA 108:12227--12232.

\bibitem[{Krebs et~al.(2001)Krebs, Boonstra, Boutin, and
  Sinclair}]{Krebsetal2001}
Krebs, C.~J., R.~Boonstra, S.~Boutin, and A.~R.~E. Sinclair. 2001.
\newblock What drives the 10-year cycle of snowshoe hares?
\newblock BioScience 51(1):25--35.

\bibitem[{Krebs and Boutin(1985)}]{Krebsetal1985}
Krebs, C.~J., and S.~Boutin. 1985.
\newblock A natural feeding experiment on a declining snowshoe hare population.
\newblock Oecologia 70:194--197.

\bibitem[{Krebs et~al.(1995)Krebs, Boutin, Sinclair, Smith, Dale, Martin, and
  Turkington}]{Krebsetal1995}
Krebs, C.~J., S.~Boutin, A.~R.~E. Sinclair, J.~N.~M. Smith, M.~R.~T. Dale,
  K.~Martin, and R.~Turkington. 1995.
\newblock Impact of food and predation on the snowshoe hare cycle.
\newblock Science 269:1112--1115.

\bibitem[{MacLulich(1937)}]{MacLulich1937}
MacLulich, D.~A. 1937.
\newblock Numbers of the varying hare.
\newblock \emph{In} University of Toronto Sutides, Biological Series 43.
  Univeristy of Toronto.

\bibitem[{Murray(2012)}]{Murray2002}
Murray, J.~D. 2012.
\newblock Mathematical Biology: I. An Introduction.
\newblock Springer.

\bibitem[{Odum(1953)}]{Odum1953}
Odum, E. 1953.
\newblock Fundamentals of Ecology.
\newblock Saunders, Philadelphia.

\bibitem[{Rosenzweig and MacArthur(1963)}]{RosenzweigMacArthur1963}
Rosenzweig, M.~L., and R.~H. MacArthur. 1963.
\newblock Graphical representation and stability conditions of predator-prey
  interactions.
\newblock Am. Nat. 97:209--223.

\bibitem[{Saint-Janin et~al.(2014)Saint-Janin, Hugueny, Aubry, Fouchet,
  Gimenez, and Pointer}]{SaintJanin2014}
Saint-Janin, H., B.~Hugueny, P.~Aubry, D.~Fouchet, O.~Gimenez, and D.~Pointer.
  2014.
\newblock Accounting for sampling error when inferring population synchrony
  from time-series data: A bayesian state-space modelling approach with
  applications.
\newblock Plos One 9(1):1--12.

\bibitem[{Santos et~al.(2014)Santos, Cabella, and Martinez}]{Santos2014}
Santos, L.~S., B.~C.~T. Cabella, and A.~S. Martinez. 2014.
\newblock {Generalized Allee effect model}.
\newblock Theor. Biosci. .

\bibitem[{Schneider et~al.(2012)Schneider, Scheu, and
  Brose}]{Schneideretal2012}
Schneider, F.~D., S.~Scheu, and U.~Brose. 2012.
\newblock {Body mass constraints on feeding rates determine the consequences of
  predator loss.}
\newblock Ecol. Lett. 15:436--43.

\bibitem[{Smith(1983)}]{Smith1983}
Smith, C.~H. 1983.
\newblock Spatial trends in canadian snowshoe hare,{\it lepus americanus},
  population cycles.
\newblock Can. Field Nat. 97:151--160.

\bibitem[{Snyder et~al.(2005)Snyder, Chang, and Prasad}]{Snyderetal2005}
Snyder, W.~E., G.~C. Chang, and R.~P. Prasad. 2005.
\newblock Ecology of Predator-Prey Interactions, chap. Conservation Biological
  Control: Biodiveristy Influences the Effectiveness of Predators, pages 324--.
\newblock Oxford University Press.

\bibitem[{Sovell and Holmes(1996)}]{SovellHolmes1996}
Sovell, J.~R., and J.~C. Holmes. 1996.
\newblock Efficacy of ivermectin against nematodes infecting field ppopulation
  of snowshoe hares (pepus americanas) in yukon, canada.
\newblock J. Wildlife Dis. .

\bibitem[{Stenseth(2007)}]{Stenseth2007}
Stenseth, N.~C. 2007.
\newblock {Canadian hare-lynx dynamics and climate variation: need for further
  interdisciplinary work on the interface between ecology and climate}.
\newblock Clim. Res. 34:91--92.

\bibitem[{Vandermeer(2004)}]{Vandermeer2004}
Vandermeer, J. 2004.
\newblock {Coupled oscillations in food webs: balancing competition and
  mutualism in simple ecological models.}
\newblock Am. Nat. 163:857--867.

\bibitem[{Weitz and Levin(2006)}]{WeitzLevin2006}
Weitz, J.~S., and S.~A. Levin. 2006.
\newblock {Size and scaling of predator-prey dynamics}.
\newblock Ecol. Lett. 9:548--557.

\bibitem[{Yan et~al.(2013)Yan, Stenseth, Krebs, and Zhang}]{Yanetal2013}
Yan, C., N.~C. Stenseth, C.~J. Krebs, and Z.~Zhang. 2013.
\newblock {Linking climate change to population cycles of hares and lynx.}
\newblock Glob. Change Biol. 19:3263--3271.

\end{thebibliography}

\end{document}